\documentclass[conference,compsoc]{IEEEtran}
\IEEEoverridecommandlockouts

\usepackage{cite}
\usepackage{amsmath,amssymb,amsfonts}
\usepackage{algorithmic}
\usepackage{graphicx}
\usepackage{textcomp}
\usepackage{xcolor}
\usepackage{listings}
\usepackage{arydshln}

\newcommand{\name}{{\tt ES-Fuzz}}
\newcommand{\nameS}{{\tt ES-Fuzz} }

\lstdefinestyle{mystyle}{
  frame=single,
  rulecolor=\color{black},
  commentstyle=\color{blue},
  keywordstyle=\bfseries\color{green!40!black},
  numberstyle=\color{black},
  stringstyle=\color{orange},
  basicstyle=\ttfamily\footnotesize,
  breakatwhitespace=false,
  breaklines=true,
  captionpos=b,
  numbers=left,
  numbersep=10pt,
  showspaces=false,
  showstringspaces=false,
  showtabs=false,
  tabsize=2,
}
\begin{document}

\title{ES-Fuzz: Adaptive Modeling of MMIO
Data Chunks for Firmware Fuzzing}

\author{
  \IEEEauthorblockN{Wei-Lun Huang}
  \IEEEauthorblockA{
  \textit{Computer Science and Engineering}\\
  \textit{University of Michigan}\\
  Ann Arbor, MI, USA\\
  weilunh@umich.edu}
  \and
  \IEEEauthorblockN{Kang G. Shin}
  \IEEEauthorblockA{
  \textit{Computer Science and Engineering}\\
  \textit{University of Michigan}\\
  Ann Arbor, MI, USA\\
  kgshin@umich.edu}
}

\maketitle

\begin{abstract}
Fuzzing has been widely used for testing
embedded-system firmware in a fully rehosted
environment without real peripherals,
native system supports, or access to 
the source code and specifications.
Some fuzzers emulate the MMIO behavior of
missing peripherals based on the firmware binaries 
to boost code coverage.
They emulate each individual MMIO read in the 
firmware with a fixed model.
We find this ineffective when multiple MMIO reads
collectively retrieve a data chunk, 
which greatly impedes the coverage growth.

We propose \nameS to overcome this coverage bottleneck
by {\em adaptively modeling MMIO data chunks}.
\nameS runs alongside a given fuzzer and 
starts a new run when the fuzzer's coverage stagnates. 
In each run, it analyzes a high-coverage test case to
infer new MMIO data-chunk models that unlock additional
execution paths.
We have implemented \nameS on Fuzzware and
evaluated it on 24 popular firmware binaries.
\nameS boosts Fuzzware's coverage by up to $68\%$
in 11---and triggers additional bugs in 5---of them 
without degrading the coverage in the remainder.
Its models describe a wide range of MMIO data chunks
and the firmware's use of each across various contexts.
\end{abstract}


\section{Introduction}
\label{sec:intro}

Embedded systems have drawn significant
attention from the security community since
they are ubiquitous, collect sensitive data,
and help users/applications make critical decisions.
However, it remains challenging to detect
vulnerabilities in their firmware at scale
due to the diversity of their platforms and purposes.
Fuzzing has been widely used to meet this challenge.
A fuzzer can test the firmware in a fully
rehosted environment by emulating the intended
microcontroller (MCU) and peripherals.
This, known as {\em rehosting-based firmware fuzzing}
\cite{pretender, p2im, laelaps, dice, jetset, uEmu,
fuzzware, sEmu, emberio, icicle, hoedur, safirefuzz,
splits, aim, multifuzz, aidfuzzer, gdma, fido, khost},
has been a popular choice for testing
embedded systems.

A fuzzer detects firmware crashes and hangs by
running the firmware with randomly generated inputs.
Its performance is typically assessed with
{\em code coverage}---the amount of firmware code
reached by the inputs.
To achieve decent code coverage,
a rehosting-based firmware fuzzer must emulate
peripheral responses to the firmware's requests.
Most existing fuzzers meet this requirement by
emulating the memory-mapped I/O (MMIO) behavior
of peripherals \cite{pretender, p2im, laelaps,
jetset, uEmu, fuzzware, sEmu, hoedur, emberio,
splits, multifuzz, khost}.
To this end, some fuzzers model MMIO data values
and configure their firmware emulators to return
these values on subsequent MMIO reads.
These fuzzers infer the firmware's {\em MMIO models}
from real-world executions \cite{pretender},
code patterns \cite{p2im},
dynamic symbolic executions (DSEs)
\cite{jetset, uEmu, fuzzware}, or
peripheral specifications \cite{sEmu}.

Using existing MMIO models, a fuzzer can emulate
MMIO reads in the firmware under test that
control peripherals, get peripheral status,
or retrieve simple data.
The fuzzer can thus progress beyond the firmware's
control- and status-intensive startup code and
reach its data-intensive application code.
The fuzzer may further cover the bare bones of
the application code, but is unlikely to cover more.
This coverage bottleneck arises because the fuzzer
uses fixed MMIO models that do not capture
the relationships among data values retrieved from
the same MMIO register.

For two reasons, such models are poor in describing MMIO 
reads in embedded-system firmware that collectively 
retrieve a data chunk.
First, they do not distinguish multiple
reads from the same MMIO register, thus limiting
their ability to represent a data chunk in which
different MMIO data follow different syntactic rules.
Second, they do not accept updates
during fuzzing after deployment, thus limiting
their ability to represent a data chunk that
the firmware processes in a context-sensitive manner.
Some fuzzers for other programs (e.g., device drivers
and virtual devices) resolve similar issues by
modeling input data chunks for each program under test.
However, they require manual annotations
\cite{ijon, videzzo}, native OS support
\cite{devfuzz, truman}, the program's source code
\cite{videzzo, truman}, or the knowledge of
hypervisor designs \cite{morphuzz}.
So, they do not fit the common settings
of firmware fuzzing.

\begin{figure*}[hbtp]
\centerline{
\includegraphics[width=0.95\textwidth]
{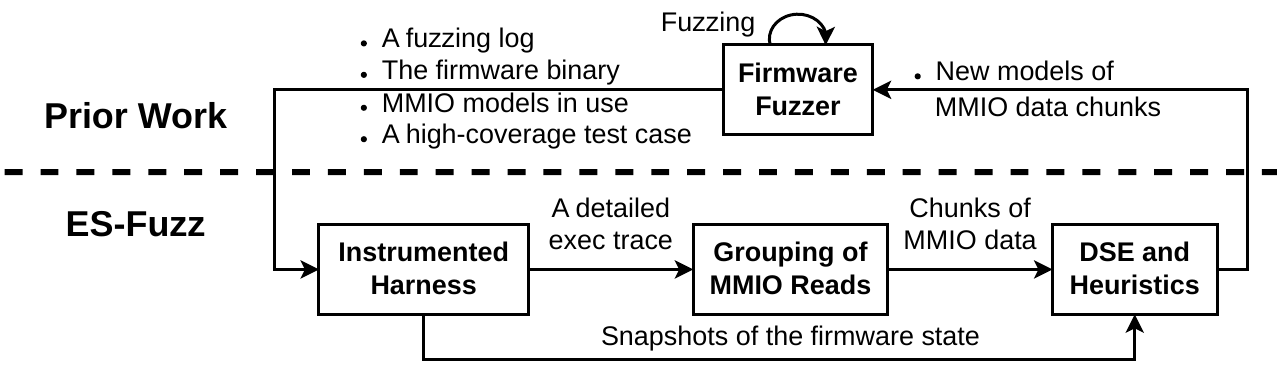}}
\caption{\nameS and a given firmware fuzzer work
together to improve the fuzzing coverage.}
\label{fig:workflow}
\end{figure*}

We propose \nameS to address the above issue by
adaptively modeling MMIO data chunks received by
the firmware in binary-only rehosting-based
firmware fuzzing.
\nameS runs alongside a given firmware fuzzer
and starts a new run whenever the fuzzer's
coverage stagnates.
In each run, it analyzes a high-coverage test case
to generate MMIO data-chunk models that are likely to
unlock additional coverage.
The test case takes the form of a detailed
execution trace produced by an instrumented
copy of the fuzzing harness (Sec.~\ref{sec:harness}).
The models will be used in the subsequent firmware
executions during fuzzing (Sec.~\ref{sec:format}).

\name's trace-guided model generation addresses
two key challenges in building adaptive models of
MMIO data chunks for firmware fuzzing
(Sec.~\ref{sec:cluster} and~\ref{sec:heuristics}).
First, each model must capture the relationships among
the values read from the same MMIO register and remain
efficiently inferable from an execution trace.
To this end, \nameS infers the model from a minimal
group of MMIO reads in the trace that, together,
retrieve a complete data chunk from the register.
These reads are grouped by when and where the firmware
consumes their returned values as a chunk, since
most embedded systems process an MMIO data chunk
only after its full retrieval.
Second, a data chunk may require distinct models across
firmware contexts when the relationships among
its constituent MMIO data are context-sensitive.
\nameS does not enumerate all contexts
or generate all models in a single run.
It favors efficient model generation,
and firmware fuzzers favor simple models.
So, it provides the given fuzzer only with
the models to overcome the coverage bottlenecks
revealed by each run's trace.

Fig.~\ref{fig:workflow} illustrates the interactions
between \nameS and the given firmware fuzzer, which is
Fuzzware \cite{fuzzware} in our implementation.
Fuzzware represents the state of the art (SOTA)
in MMIO modeling for firmware fuzzing and
serves as an MMIO modeling extension for later
fuzzers that address orthogonal research problems
\cite{hoedur, aidfuzzer, gdma, khost, fido}.
We have evaluated both \nameS and vanilla Fuzzware on
firmware benchmarks commonly used in prior work.
\nameS is shown to improve Fuzzware's coverage by
up to $68\%$ and discover additional firmware vulnerabilities.
Its MMIO models capture the firmware's expectations
for a wide range of MMIO data chunks across
diverse data-usage contexts.
By evaluating our DSE heuristics for mitigating
path explosion, we have demonstrated that vanilla DSE
\cite{laelaps, devfuzz} cannot construct most of
\name's MMIO models within reasonable CPU and memory budgets.


\section{Background}
\label{sec:background}

\subsection{Firmware--Peripheral Interactions}
\label{sec:interaction}

In embedded systems, firmware continuously interacts
with the intended peripherals during execution and
adapts its control flow based on their responses.
Thus, a test case (i.e., an input) in firmware fuzzing
refers to the collection of such interactions during
a firmware execution.
These interactions are primarily realized through
memory-mapped I/O, interrupt requests,
and direct memory access.

\textbf{Memory-mapped I/O (MMIO).}
Firmware interacts with peripherals through MMIO
by accessing designated memory regions where
each address represents an {\em MMIO register}.
Each peripheral maps its registers and on-device memory
to a pre-assigned set of MMIO registers for
the firmware to read and write.
MMIO registers commonly fall into
three categories \cite{p2im}:
\begin{itemize}
  \item \underline{Control Register (CR)}:
  The firmware controls a peripheral
  by writing to the peripheral's CR.
  \item \underline{Status Register (SR)}:
  A peripheral reports its current status by
  updating its SR, which the firmware reads
  to determine its control flow.
  \item \underline{Data Register (DR)}:
  The firmware can send a complex command to an actuator
  by writing command bytes sequentially to the actuator's DR.
  A sensor can place its latest reading in its DR
  for the firmware to read.
  Peripherals like serial communication interfaces
  have DRs that operate in both directions.
\end{itemize}
A CR/SR value typically consists of independent bit fields,
whereas a DR value may be multi-byte wide and
depend on the previous values in that DR.
Thus, DR reads are generally harder to emulate
in rehosting-based firmware fuzzing than CR or SR reads.
Bit fields of different functions (e.g., control bits
and status bits) can coexist in an MMIO register \cite{p2im}.

A lasting research problem in rehosting-based
firmware fuzzing is how to classify
the MMIO registers accessed during execution.
A baseline solution relies on the CMSIS-SVD file
\cite{cmsis, pretender} of the target MCU.
SOTA fuzzers \cite{p2im, uEmu, fuzzware} adopt
alternatives to accurately identify DRs.
They will emulate DR reads to return arbitrary values
because DR reads are hard to emulate and
DRs are assumed to be under adversarial control
when the firmware is under attack.

\textbf{Interrupt Request (IRQ).}
Firmware receives IRQs from peripherals
for asynchronous interactions.
It maintains an interrupt vector table (IVT) that
maps each IRQ type to a function address.
This function is an {\em interrupt service routine}
(ISR) that handles all incoming IRQs of this type.
Upon receiving an IRQ, the firmware consults the IVT and
context-switches to the appropriate ISR based on the IRQ type.
In this paper, we will design policies not for emulating IRQs
\cite{pretender, aim, sEmu, aidfuzzer, fido} but
for emulating DR reads in IRQ contexts (i.e., in ISRs),
whose returned values are often consumed
by firmware outside the contexts \cite{uEmu}.

\textbf{Direct Memory Access (DMA).}
DMA allows a peripheral to transfer data
to/from firmware memory without CPU intervention.
Before the transfer, the firmware configures
the DMA controller on its MCU with the data source
(e.g., a peripheral's DR), destination (e.g.,
a global \texttt{uint8} array), and size.
When the data are ready, the DMA controller moves
them from the source to the destination independently
of the MCU cores.
However, DMA emulation \cite{dice, sEmu, gdma} is
outside the scope of this paper.

\subsection{American Fuzzy Lop (AFL)}
\label{sec:fuzzer}

AFL \cite{afl, aflpp} is a popular coverage-guided
fuzzer for general-purpose software.
It can test software by running multiple fuzzing instances
in parallel and coordinating them via the main instance
to speed up coverage growth.
Each instance may adopt a different policy for
mutating and scheduling test cases.
If AFL cannot run the software in the host environment,
it emulates the software with QEMU \cite{qemu} or
Unicorn \cite{unicorn}.
Users may provide a {\em fuzzing harness} to
specify the emulation setup and the software's
interpretation of a test case.
Most rehosting-based firmware fuzzers mutate and schedule
test cases with AFL, and translate each test case into
a sequence of firmware--peripheral interactions
via a customized harness.

\subsection{Dynamic Symbolic Execution (DSE)}
\label{sec:symexec}

Symbolic execution is a program-analysis technique
that assigns symbols rather than concrete values
to program inputs and executes the program along
all possible paths.
Each executed instruction expresses its result
in terms of the input symbols.
Each path then leads to
a {\em path constraint}---the logical conjunction
of all branch conditions along that path.
An SMT solver such as Z3 \cite{z3} can find
{\em satisfying assignments} for a satisfiable
path constraint, which are concrete program inputs
that satisfy the constraint.
An unsatisfiable path constraint indicates
that the program never takes the path when
running with concrete inputs.
The solver can examine a path constraint
together with constraints that encode events of interest.
Each satisfying assignment in this case
is an input that drives the program along that path
and triggers those events.

It is generally infeasible to execute an entire program
symbolically, as the number of execution paths grows
exponentially with the program size.
DSE (i.e., concolic execution) mitigates
this path explosion by running most of the program
with concrete inputs and only the code of interest
with symbolic inputs.
Prior works on MMIO emulation for firmware fuzzing
apply DSE to the firmware code surrounding MMIO reads
\cite{laelaps, jetset, uEmu, fuzzware, sEmu}.
Some of them further prune or prioritize
symbolic firmware states using heuristics,
as vanilla DSE alone is insufficient to contain
path explosion \cite{devfuzz, laelaps}.
Sec.~\ref{sec:efficiency}
and Appendix~\ref{sec:devfuzz}
will cover this issue.

\section{Overview of \name}
\label{sec:overview}

This section provides an overview of \name.
Sec.~\ref{sec:model} introduces
the MMIO models generated by \name.
Sec.~\ref{sec:motivation} motivates their use
in rehosting-based firmware fuzzing through
a real-world example.
Sec.~\ref{sec:workflow} describes
how \nameS cooperates with a given firmware fuzzer
to generate and deploy these models during fuzzing.

\subsection{Adaptive Models of MMIO Data Chunks}
\label{sec:model}

In rehosting-based firmware fuzzing, the firmware
interacts with emulated peripherals during each execution
according to a fuzzer-generated schedule.
A typical fuzzer produces each schedule (i.e.,
each test case) in two steps.
First, a coverage-guided fuzzing engine (e.g., AFL)
generates a random byte string.
Then, a fuzzing harness translates this byte string into
a schedule based on the firmware's peripheral models
(e.g., MMIO models).
Some fuzzers, such as Hoedur \cite{hoedur}
and MultiFuzz \cite{multifuzz}, aim to
improve the first step.
Others, such as Fuzzware \cite{fuzzware}
and \name, aim to improve the second.
Firmware treats MMIO accesses as memory accesses, so
a firmware fuzzer's \textbf{only} uncertainty in
emulating MMIO accesses lies in the values returned
by MMIO reads.
MMIO models used in the second step
primarily describe these returned values
\cite{pretender, p2im, jetset, uEmu, fuzzware, sEmu}.
The fuzzer assigns each MMIO read with a model
inferred from, say, the firmware binary.
The read returns a value prescribed by the model
upon emulation, consuming random bytes generated by
the fuzzing engine if the model admits multiple values.

\begin{lstlisting}[
language=bash,
numbers=none,
xleftmargin=1mm,
xrightmargin=1mm,
caption={Fuzzware's MMIO model},
label={lst:fwmods}]
# one random byte consumed per modeled MMIO read
pc_000813cc_mmio_400e0818:
  access_size: 0x4
  mask: 0xff # entropy: 1 byte
  left_shift: 0x0
\end{lstlisting}
\begin{lstlisting}[
language=bash,
numbers=none,
xleftmargin=1mm,
xrightmargin=1mm,
caption={\name's MMIO model},
label={lst:efmods}]
# one random byte consumed per model selection
irq_18_mmio_400e0818:
  0x00: undef # fall back on Fuzzware's model
  # "steer,\n", "motor,\n"
  0x01: [0x73,0x74,0x65,0x65,0x72,0x2c,0x0a] 
  0x02: [0x6d,0x6f,0x74,0x6f,0x72,0x2c,0x0a] ...
  # "motor,-9\n", "steer,6\n", "motor,214\n"
  0x0e: [0x6d, ..., 0x2c,0x2d,0x39,0x0a]
  0x0f: [0x73, ..., 0x2c,0x36,0x0a]
  0x10: [0x6d, ..., 0x2c,0x32,0x31,0x34,0x0a]
\end{lstlisting}

Fuzzware and \nameS generate the MMIO models
shown in Listings~\ref{lst:fwmods} and~\ref{lst:efmods}
when testing the Steering Control firmware
from P2IM \cite{p2im}.
Both models admit multiple values and describe
firmware reads from the MMIO register at 0x400e0818.
Fuzzware names its model after the common instruction
address of these reads; \nameS names its model after
the common type of their IRQ contexts.
Fuzzware models each of these reads independently
by consuming a random byte 0x$XY$ and
returning 0x000000$XY$ for that read.
\nameS models these reads \textbf{collectively} with
16 single-valued models of their returned data chunk.
It consumes one random byte $Z$ and computes
$z = Z \bmod 17$ to select the $z$-th single-valued model.
The subsequent reads will retrieve the data chunk
specified by that model.
Moreover, \name's model is \textbf{adaptive}.
Its later single-valued models are generated
in its later runs and capture richer syntax/semantics
of the modeled data chunk.

\subsection{A Motivating Example}
\label{sec:motivation}

\begin{lstlisting}[language=C,
morekeywords={true, false},
emph={Servo, String},
emphstyle=\bfseries\color{red!50!black},
xleftmargin=7mm,
xrightmargin=3mm,
caption={Steering Control from P2IM (simplified)},
label={lst:strctl}]
String cmd, val; // 0 <= int(val) <= 180
Servo steering, throttle;
while (Serial.available() > 0) {
  cmd = Serial.readStringUntil(',');
  val = Serial.readStringUntil('\n');
  if (cmd == "steer")
    steering.write(val.toInt());
  else if (cmd == "motor")
    throttle.write(val.toInt());
}
\end{lstlisting}

Most rehosting-based firmware fuzzers
model each MMIO read independently and
do not update the MMIO models they generate.
Such models fail to capture the data chunks
retrieved via multiple MMIO reads and
processed by the firmware in a context-sensitive manner,
thereby limiting a fuzzer’s coverage of
data-intensive application code.
We henceforth reuse the firmware and MMIO models
from Section~\ref{sec:model} to illustrate
this limitation and motivate the need
for \name’s MMIO models in firmware fuzzing.

The firmware periodically executes the loop
in Listing~\ref{lst:strctl} to parse newly received
commands and adjust its servo motors.
It reads the commands character by character from
a data register (UART RHR) at 0x400e0818.
Each well-formed command specifies a motor name \texttt{cmd}
and a parameter \texttt{val}, separated by a comma:
\texttt{cmd} is ``steer'' or ``motor'';
\texttt{val} evaluates to an integer in the range 0--180.
For a well-formed command, the firmware writes
\texttt{val} to the motor named \texttt{cmd}.
\texttt{val < 0} and \texttt{val > 180} trigger
different error-handling branches in the firmware.

In the above example, different characters
in a command (letters, digits, and the comma) obey
distinct syntactic rules.
The model in Listing~\ref{lst:fwmods} treats every
UART RHR read identically and cannot distinguish,
for instance, a letter read from a digit read.
In contrast, the model in Listing~\ref{lst:efmods}
characterizes each read according to its role in
retrieving a command.
The example also shows that
\texttt{val}'s semantics depend on \texttt{cmd}.
If a fuzzer emulates all the UART RHR reads with
a single fixed model, that model must account for
all the \texttt{cmd}-dependent \texttt{val} semantics,
making itself too generic to use (e.g., the model
in Listing~\ref{lst:fwmods}) or too complex to build
in one shot.
In contrast, the model in Listing~\ref{lst:efmods}
evolves during fuzzing to capture more command semantics
through more single-valued models.

By comparing Fuzzware's and \name's models, we
highlight a limitation of SOTA MMIO models
for rehosting-based firmware fuzzing.
Existing models cannot properly describe MMIO reads
that collectively retrieve a data chunk.
As a result, fuzzers will struggle to cover much of
the application code in embedded-system firmware.
To overcome this coverage bottleneck, they should emulate
such MMIO reads using adaptive models of MMIO data chunks.

\subsection{System Workflow}
\label{sec:workflow}

\nameS generates the MMIO models introduced
in Sec.~\ref{sec:model} to address the limitation
illustrated in Sec.~\ref{sec:motivation}.
As depicted in Fig.~\ref{fig:workflow},
it runs alongside a given rehosting-based firmware fuzzer.
It starts a new run when the fuzzer fails to deploy
more MMIO models or cover more basic blocks (BBs)
in the firmware under test for a sustained period.
In each run, it analyzes a high-coverage test case
to infer MMIO data-chunk models that guide
the fuzzer toward previously unexplored BBs.
Compared to existing firmware fuzzers, \nameS
progressively builds and deploys MMIO models that
capture richer syntax and semantics of MMIO data chunks.
Compared to non-firmware fuzzers that model
structured program inputs, \nameS adheres to
the common constraints of firmware fuzzing:
no manual annotations, native OS support, or source code.

In each run, \nameS analyzes a test case that covers
the most (at least one) of the BBs not covered by
any previously analyzed test cases.
It cancels the run if no such test case exists.
To identify this test case, it queries the given
firmware fuzzer for all the MMIO models in use,
test cases, and lists of BBs covered by each test case.
Most existing fuzzers produce such lists on demand
for debugging and development purposes.
Based on the identified test case, \nameS then builds
new MMIO data-chunk models, which the fuzzer deploys
in subsequent firmware executions to overcome
its current coverage bottleneck.
The iterative interactions between \nameS
and the fuzzer progressively improve the coverage.
So far, we have described \name's workflow by
treating each run as a black box.
Next, we will cover the details of a single run.

\section{System Design and Implementation}
\label{sec:design}


This section details \name's design and implementation.
Fig.~\ref{fig:workflow} and Sec.~\ref{sec:overview}
have outlined its interactions with
a given firmware fuzzer between runs.
We now focus on its three steps in each run to infer
new MMIO data-chunk models from a selected test case.

First, \nameS obtains a detailed trace of the firmware
under test executing the selected test case
(see Sec.~\ref{sec:harness}).
The trace records every executed
BB and MMIO access in chronological order.
It further highlights MMIO reads that retrieve
parts of a data chunk and records the firmware's
subsequent use of their MMIO data.
The trace is generated by an instrumented copy of
the given fuzzing harness, which we construct by
manually instrumenting the harness \textbf{once}
in a fuzzer-dependent but firmware-agnostic manner.

Second, \nameS groups the highlighted MMIO reads
in the trace by when during execution and
where in RAM the firmware uses their MMIO data
(see Sec.~\ref{sec:cluster}).
It then treats each group as collective retrieval
of a complete MMIO data chunk and infers
the modeling priority of these chunks from
statistics collected during grouping.
Each group is optionally matched against
constants in the firmware binary.

Lastly, \nameS attempts to model the values returned
by each group of MMIO reads through DSE and heuristics
(see Sec.~\ref{sec:heuristics}).
The DSE per group targets the firmware code
where the MMIO reads occur and
where the retrieved data chunk is used.
When it runs the MMIO-read code, it follows the path
recorded in the detailed execution trace.
When it runs the data-use code, it prioritizes
and prunes symbolic firmware states with heuristics.
Each DSE instance begins with an appropriate snapshot
of the firmware state taken by the instrumented harness.

If \nameS successfully builds new MMIO data-chunk models
by taking the above steps, the fuzzer adopts them
(see Sec.~\ref{sec:format}).
These are expected to fit the modeled chunks
better than the fuzzer's native models.
Sec.~\ref{sec:implementation} describes
our implementation of \name.

\subsection{The Instrumented Harness}
\label{sec:harness}

\begin{lstlisting}[
language=bash,
numbers=none,
xleftmargin=1mm,
xrightmargin=1mm,
caption={Some events in a detailed execution trace},
label={lst:trace}]
# four BBs
BB 17ed ADDR 813cc ISR  17eb 80718 18 # UART
BB 1fc3 ADDR 813cc ISR  1fc1 80718 18 # UART
BB 1feb ADDR 82654 FUNC 1fea 825a8  0 # strcpy
BB 665f ADDR 824de FUNC 665b 8236c  0 # strcmp
# two MMIO reads (taint sources)
BB 17ed PC 813cc MMIO 400e0818 VAL 0000006d # "m"
BB 1fc3 PC 813cc MMIO 400e0818 VAL 0000006f # "o"
# three taint sinks: write / read / non-memory
BB 1feb PC 82658 RAM  20071271 VAL 6f
  WRITE                        TO 0 20071270 2
  MMIO   1fc3 813cc 400e0818 FROM 0 20070c08 2
BB 665f PC 824e4 RAM  20071270 VAL 6f746f6d
  READ                       FROM 0 20071270 5
  MMIO   17ed 813cc 400e0818 FROM 0 20071270 5
  MMIO   1fc3 813cc 400e0818 FROM 0 20071270 5
  ...
BB 665f PC 82502
  MMIO   17ed 813cc 400e0818 FROM 0 20071270 5
  MMIO   1fc3 813cc 400e0818 FROM 0 20071270 5
  ...
  CONST 84b17    74    84b13 # "t" in "steer"
\end{lstlisting}

In each run, \nameS diagnoses the coverage bottleneck
of the underlying firmware fuzzer using the test case
described in Sec.~\ref{sec:workflow}.
This test case is presumed to cover the largest
portion of the firmware's application code
not yet analyzed by \name.
\nameS will analyze MMIO reads in this code that
retrieve parts of a data chunk, as some of them
may have caused the bottleneck by returning
undesired values.
To identify such reads, \nameS needs a detailed
trace of the firmware executing the test case.
We manually instrument a copy of the given
fuzzing harness to generate this trace on demand.
\nameS will reuse this instrumented harness across
runs and firmware under test,
unless the fuzzer is replaced.

While running a test case, the harness records
the visited BBs, the executed MMIO accesses,
and the results of a dynamic taint analysis (DTA)
in a chronological trace.
Listing~\ref{lst:trace} is an excerpt from
such a trace when \nameS tests the Steering Control
in Sec.~\ref{sec:overview}.
The harness encodes each number in hexadecimal
and timestamps each event with the dynamic BB count
and the program-counter (PC) value.
It records the ISR---or function outside
IRQ contexts---where each BB is visited
as the {\em BB context}, which applies to
all events within that BB and is denoted by
the ISR/function entry timestamp and
the associated IRQ type.

The harness tracks the firmware's use of MMIO data
through an instruction-level DTA that treats values
returned by MMIO reads as sources, except those
excluded by the fuzzer-specific heuristics
in Appendix~\ref{sec:taintless}.
Later, \nameS will use the DTA results to group
MMIO reads that collectively retrieve a data chunk.
The harness taints each CPU register or RAM location
with sources that determine the current value therein
and maintains, in real time, a list of tainted RAM buffers.
Each buffer ${<}F, A, S{>}$ contains as many contiguous
RAM bytes tainted by sources from the same MMIO registers
as possible.
$F$ is the entry timestamp of the allocating
ISR or function (0 for global buffers).
$A$ is the start address:
0x20070c08 for the UART buffer and 0x20071270 for
an Arduino String in Listing~\ref{lst:trace}.
$S$ is the current size.
Each taint during the DTA is labeled with the buffer
from which it was loaded last time, when applicable.
Taints within a call frame are cleared when
the firmware exits the corresponding ISR or function.
The DTA ignores branch instructions for efficiency.

We define sinks in the above DTA as instructions
tainted by an MMIO data outside {\em the MMIO read's
BB context and its caller and callee chains}.
This definition follows from the observation that
if an MMIO read retrieves part of a data chunk,
the returned value is mostly consumed outside
the read's context \cite{uEmu}.
The sinks exclude data movements such as
the \texttt{mov} instructions and
RAM accesses to the call stack.
As shown in Listing~\ref{lst:trace}, for each sink,
the harness records its taint sources and
the RAM buffers through which its taints propagate.
The sources reveal MMIO data within the same chunk,
while the buffers capture distinct contexts in which
the firmware consumes this chunk.

The harness tracks the firmware's use of
constant strings to speed up \name's DSE later.
It extends sources in the above DTA to
include bytes of the constant strings
loaded from the firmware binary into memory.
This enables the DTA to record string bytes
that the firmware uses alongside MMIO data.
For example, the last line in Listing~\ref{lst:trace}
refers to the byte ``t'' (i.e., 0x74) at address 0x84b13
in the string ``steer'' that ends at 0x84b17.

\begin{figure*}[hbtp]
\centerline{
\includegraphics[width=0.95\textwidth]
{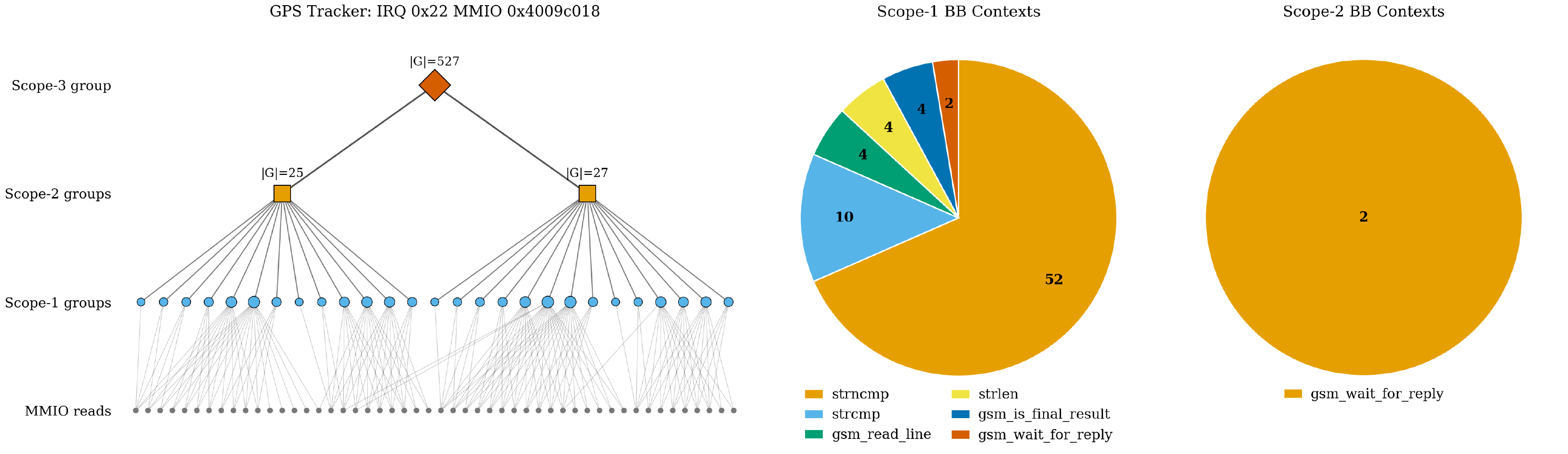}}
\caption{\nameS groups MMIO reads in the given
firmware trace based on the firmware's co-usage
of them in BB contexts at three scopes.}
\label{fig:clusters}
\end{figure*}

The harness provides snapshots of the firmware state
upon request to ensure that \name's DSE starts with
an appropriate state.
Each request specifies a test case and a set of 
timestamps (dynamic BB counts and PC values).
Once requested, the harness runs the firmware with
the specified test case and snapshots CPU registers
and the allocated memory at each specified time.

\subsection{Grouping of MMIO Reads}
\label{sec:cluster}

\nameS now has a detailed trace of the firmware
under test executing the selected test case.
In this trace, the DTA in Sec.~\ref{sec:harness}
records the firmware's uses (sinks) of MMIO data
(sources) outside the MMIO-read contexts.
\nameS treats each sink as the use of an MMIO data chunk
and each associated source as part of that chunk.
It will group the taint sources in the trace so that
each group is presumed to form a complete chunk.
Then, it will infer the modeling priority of
these chunks and match each chunk with literals (e.g.,
strings) against which the firmware compares the chunk.

\nameS groups these sources hierarchically
based on the coexistence of their sinks
at three scopes:
(1) within a BB context (ISR/function),
(2) along a call chain, and (3) across call chains.
This design follows from our observation that
embedded-system firmware tends to use the MMIO data
targeted by \nameS collectively after these data
form a meaningful chunk.
At Scope 1, \nameS considers only BB contexts
that contain at least one sink.
For each BB context, \nameS groups its sinks
associated with the same tainted RAM buffer,
converts each sink group into a group of
the associated sources, and merges overlapping
source groups.
At Scopes 2 and 3, \nameS will merge the source groups
obtained at Scope 1 and record the input and output
groups of each merge.

To prepare for Scope 2, \nameS infers
the call chains along which the firmware
uses the targeted MMIO data across BB contexts.
\nameS considers only the Scope-1 BB contexts that
represent function invocations and defines
the {\em lifetime} of each context as the interval
from its function entry to its last sink.
Then, \nameS constructs trees over these contexts
by treating a context $C_1$ as an ancestor of $C_2$
if $C_1$'s lifetime contains $C_2$'s.
In these {\em call trees}, each root-to-node path
represents a call chain of \name's interest:
each caller along the path may use MMIO-dependent
results returned by its callees.
Each Scope-1 BB context that represents an ISR invocation
is treated as a single-node call tree because
context switches isolate it from other BB contexts.

\nameS obtains Scope-2 source groups by traversing
each call tree in post-order (i.e., bottom-up).
It merges the Scope-1 groups of each visited tree node
with overlapping Scope-2 groups from the node's children.
The node's Scope-2 groups are then all its Scope-1 groups
and the remaining Scope-2 groups from children
after such merges.
\nameS obtains Scope-3 groups by merging
overlapping Scope-2 groups across all tree roots.

A Scope-1 group is expected to identify
a minimally meaningful chunk of MMIO data.
A Scope-2 group is expected to combine
semantically related smaller chunks into
a larger meaningful chunk.
A Scope-3 group is expected to capture
long-term relationships among large chunks.
After the three-stage grouping of MMIO reads,
or equivalently their returned MMIO data,
\nameS prepares the resulting groups for
its next step: DSE and heuristics.

First, \nameS partitions the reads in each group
by the IRQ types of their BB contexts and
the addresses of their MMIO registers.
This way, all reads in a group occur
in the same type of IRQ contexts
and at the same MMIO register, which
enables the DSE to apply aggressive heuristics.

Second, \nameS consolidates identical groups
and, for each resulting group, records the BB contexts
in which the group appears at each scope.
The DSE will prioritize groups that appear
in more diverse contexts (in terms of IRQ type
and ISR/function address) and, secondarily,
in more contexts overall.
To this end, it will compare these statistics
from Scope 1 to 3 and break any remaining ties by
comparing the grouped MMIO reads lexicographically.
Context diversity and multiplicity reflect
a group's importance to the firmware under test.
The tie-breaker favors groups that contain earlier and,
secondarily, fewer MMIO reads, which typically yield
more resource-efficient DSE instances.

\begin{table*}[hbtp]
\centering
\caption{Heuristics for \name's DSE to
prioritize and prune symbolic firmware states}
\label{tab:heuristics}
\begin{tabular}{|l|l|l|}
\hline
\multicolumn{1}{|c|}{\textbf{Action}} &
\multicolumn{1}{c|}{\textbf{Criterion or Condition}} &
\multicolumn{1}{c|}{\textbf{Reason}} \\
\hline
The DSE always &
(C1) with the fewest symbols in its path constraint, &
This delays path explosion. \\
\cline{2-3}
takes the state &
(C2) closest to \textbf{any} unvisited BB nearby, and &
The DSE will then reach an unvisited BB early. \\
\cline{2-3}
in the queue $\ldots$ &
(C3) waiting for the longest time. &
FIFO breaks priority ties. \\
\hline
The DSE prunes &
(C4) an MMIO read executed at the state &
The DSE can no longer efficiently emulate \\
a state when $\ldots$ &
cannot be mapped to any in the trace. &
the MMIO reads outside \name's interest. \\
\cline{2-3}
&
(C5) the state has a cycle in
its history of register contents. &
The DSE is stuck in an infinite loop \cite{uEmu}. \\
\cline{2-3}
&
(C6) the state has no symbolic RAM contents.&
The state no longer depends on the \\
&
&
MMIO reads of \name's interest. \\
\hline
\end{tabular}
\end{table*}

Lastly, \nameS matches an MMIO-read group with
a firmware string if the given trace indicates that
the firmware once used the group's MMIO data together
with the string bytes.
If the MMIO data and string bytes were ever co-used
by a memory instruction, \nameS assumes the use case
is string concatenation and skips the match.
Otherwise, it assumes the use case is a string comparison
and matches the group with the string starting from
their earliest co-used data and bytes.
The DSE instance for this group will convert each
matched string into a set of constraints on the grouped
MMIO reads to mitigate path explosion.

Fig.~\ref{fig:clusters} illustrates \name's
grouping of MMIO reads through a Scope-3 group
built by \nameS for testing the GPS Tracker firmware
from uEmu \cite{uEmu}.
This group contains 527 reads
from 171 unique Scope-2 groups.
Two of these groups are newly built at Scope 2
rather than inherited from Scope 1, containing
25 and 27 reads from 13 and 14 unique
Scope-1 groups, respectively.
Both come from BB contexts that represent
\texttt{gsm\_wait\_for\_reply} invocations,
hence exhibiting similar constituent Scope-1 group
sizes and mappings to MMIO reads.
\nameS identifies 76 occurrences of the 27 unique
Scope-1 groups, mostly during command parsing
and string comparisons.
Thus, the groups in Fig.~\ref{fig:clusters} are
matched with 21 AT-response strings from the firmware.

\subsection{DSE and Heuristics}
\label{sec:heuristics}

\nameS has grouped the MMIO reads in a firmware trace
as described in Sec.~\ref{sec:cluster} and presumed
each group to have collectively retrieved a complete
data chunk during the recorded execution.
Next, it will configure and run DSE on each group
to model the retrieved chunk value and
guide the underlying firmware fuzzer
toward previously unvisited BBs.
The DSE will run the firmware code where
the chunk is (1) retrieved and (2) used.
Such code often spans multiple functions, making it
difficult for a naïve DSE along arbitrary paths to
reach unvisited BBs within a reasonable resource budget.
\nameS will accelerate the DSE using
the given trace and our heuristics.

\subsubsection{Symbolic Execution of the Chunk Retrieval}
\label{sec:readexec}

In embedded systems, firmware that retrieves
a data chunk via multiple MMIO reads typically
validates the chunk in its usage code
rather than its retrieval code.
So, \name's DSE can execute each group
of MMIO reads along the path recorded in the given trace
without sacrificing opportunities for code coverage.
However, in practice, these reads are executed
differently depending on their BB contexts.

If these reads occur in functions
outside IRQ contexts, they are often near
the code that uses the retrieved chunk.
In such a case, the DSE executes chunk retrieval and usage
together, starting from a snapshot of the firmware state
at the first occurrence of the reads.
If these reads occur within ISRs, they are often
independent of and sometimes far from the code
that uses the retrieved chunk.
In this case, the DSE executes
chunk retrieval and usage separately.
It replays chunk retrieval by identifying ISRs that
contain these reads and executing them as many times
as recorded in the trace, in the recorded order,
and along the recorded path.
In either case, an executed MMIO read returns
a symbolic value if it is among these reads,
else the recorded value.

Before executing the identified ISRs,
the DSE on a group of MMIO reads locates
the first DTA sink in the trace whose sources
include some of the grouped reads.
This sink marks the moment $t$ at which
the firmware starts using the chunk
retrieved by these reads.
The instrumented harness in Sec.~\ref{sec:harness}
then provides the firmware-state snapshots $S_{u}$
at $t$ and $S_{r}$ at each entry to the identified ISRs
that occur before $t$.
The DSE executes these ISRs from their respective $S_{r}$
and, upon each ISR exit, transfers the remaining
symbolic RAM contents outside the call stack to $S_{u}$.
After replaying all such ISRs, it will begin executing
the chunk usage from $S_{u}$ and attempt to execute
the remaining identified ISRs at each symbolic
firmware state it reaches.
Each attempt aborts when an ISR overwrites
unused symbolic RAM contents or deviates
from the path recorded in the trace.

\subsubsection{Symbolic Execution of the Chunk Usage}
\label{sec:useexec}

The DSE on each group $G$ of MMIO reads
starts from the firmware-state snapshot prepared
in Sec.~\ref{sec:readexec} to execute the code
that uses the chunk retrieved by $G$.
To efficiently emulate the MMIO reads outside $G$,
the DSE maintains two pointers for each symbolic state:
a BB pointer and an MMIO pointer that reference
a BB and an MMIO access in the given trace, respectively.
When the DSE executes an MMIO read at a symbolic state,
these pointers map the read to its counterpart
in the trace.
The read returns a symbolic value if the counterpart
belongs to $G$, else the concrete value returned
by the counterpart.

The DSE initializes the two pointers of
its initial symbolic state as follows.
If the reads in $G$ occur in ISRs,
the BB pointer references the BB where the firmware
starts using the retrieved chunk, and
the MMIO pointer references the first MMIO access
after this BB in the trace.
Otherwise, the MMIO pointer references
the earliest read in $G$, and the BB pointer
references the corresponding BB.

As the DSE proceeds, each successor state created by
a branch is mapped to the first BB in the trace with
the same BB address and IRQ type that follows
the BB referenced by its predecessor's BB pointer.
The successor's BB pointer is updated to
reference this BB, or $\infty$ if no match exists.
Similarly, each MMIO access executed at a symbolic state
is mapped to the first MMIO access in the trace with
the same MMIO register, PC value, and IRQ type that
follows the MMIO access referenced by the state's
MMIO pointer.
If the state's BB pointer is not $\infty$, the matching
MMIO access must also follow the referenced BB.
The state's MMIO pointer is updated
to reference this access.

If the DSE treats all symbolic states equally,
it will likely suffer path explosion before reaching
any unvisited BBs.
To mitigate this problem, it maintains a priority queue
of states and prioritizes/prunes them using
the heuristics in Table~\ref{tab:heuristics}.
At first, the queue contains only the initial state.
The DSE repeatedly dequeues a state, runs with it
until a branch is reached, and enqueues
the resulting successor states.
For C2 in Table~\ref{tab:heuristics},
the DSE computes all-pairs shortest paths
in the firmware's static control-flow graph (CFG)
and considers only paths from the state's BB
to unvisited BBs in the same function or its callees.
For C5, a state's history of register contents records
the register values of its ancestor states and
resets whenever an MMIO read occurs.
For C6, the DSE zeroes the appropriate portion of
the call stack upon exiting an ISR or a function
outside IRQ contexts to maintain an accurate view
of symbolic RAM.


The DSE on a group of MMIO reads terminates
when it times out, when its queue becomes empty, or
when it reaches its first unvisited BB after
the first $P$ dequeued states
but none within the last $P$.
Whenever the DSE reaches an unvisited BB,
an SMT solver finds concrete values returned
by the MMIO reads that satisfy the path constraint.
\nameS will build a new MMIO data-chunk model
from these values.

\subsubsection{Heuristics for Scheduling DSE Instances}
\label{sec:priority}

Each group of MMIO reads has been labeled with
the common IRQ type and MMIO register of its reads
and assigned a DSE priority in Sec.~\ref{sec:cluster}.
\name's DSE then executes groups with different labels
in parallel and groups with the same label sequentially
in priority order.
If it reaches an unvisited BB while executing a group,
it skips the remaining groups with the same label.
For a group matched with firmware strings, the DSE
may speed up by forcing some reads in the group
to return bytes from a matched string.
It explores different strings in parallel and,
for each string, different read-to-byte matches
sequentially.

\subsection{Model Generation and Deployment}
\label{sec:format}

Suppose the DSE in Sec.~\ref{sec:heuristics}
reaches an unvisited BB while executing a group
of MMIO reads labeled $(I, R)$, where
$I$ and $R$ denote the common IRQ type and MMIO register
of these reads.
Then, the DSE returns the concrete
MMIO data chunk retrieved by the group.
\nameS builds an MMIO model from this chunk value.
To guide the given firmware fuzzer toward
the DSE-reached BB, the model lists the values returned
by the MMIO reads in chronological order.
A read returns an undefined value
if unconstrained by the DSE.
The fuzzer thereafter applies the model to
MMIO reads labeled $(I, R)$ in the firmware under test.

The fuzzer deploys this model in subsequent
firmware executions alongside the previous models
from \nameS for the same MMIO reads, as shown
in Fig.~\ref{fig:workflow} and Listing~\ref{lst:efmods}.
It also deploys a {\em dummy} model
as a fallback to its native models.
Before each firmware execution, or whenever
the current model is exhausted, the fuzzer
randomly selects a model.
Each modeled MMIO read then returns the next
unused value in the selected model.
If the value is undefined, the read
falls back to the fuzzer's native models.

\nameS starts its next run when the fuzzer
has not covered additional firmware BBs or
deployed new MMIO models for a while.
It tracks the BBs reached by its DSE,
the MMIO models generated, and
the matches between firmware literals and MMIO reads
explored in previous runs.
It no longer treats these BBs as unvisited,
generates these models, or explores these matches
in this run.

\subsection{Implementation}
\label{sec:implementation}




We implement \nameS for the popular ARM
Cortex-M architecture, using Fuzzware \cite{fuzzware}
(commit 28ce2dc) as the underlying firmware fuzzer
and \texttt{angr} \cite{angr} (9.2.61) as the DSE engine.
This implementation in Python, C, and Bash contains
(1) an instrumented copy of Fuzzware's harness and
(2) a program that builds MMIO data-chunk models
from a detailed firmware trace.
The two parts require $\sim$1650 and $\sim$3200 lines
of code, respectively, plus another $\sim$250 lines
for integration with Fuzzware.
Fuzzware's harness is modified to support
\name's models and to follow the same path as
its instrumented copy when executing the same test case.
To achieve the latter, it counts native firmware BBs
instead of translation blocks when scheduling
periodic firmware IRQs during execution.

Fuzzware and \nameS run concurrently on different
cores (see Sec.~\ref{sec:eval} for core counts).
Fuzzware continuously tests firmware and
records the BBs covered by each test case.
\nameS selects the test case to analyze
in each run based on these records, as
described in Sec.~\ref{sec:workflow}.
It starts a new run when such a test case exists
and Fuzzware has not covered more BBs
or deployed new MMIO models for $N=10$ minutes.
We recommend $5 \le N \le 30$ for 24-hour fuzzing such that
\nameS can provide new MMIO models frequently enough
without disrupting Fuzzware's input mutations too often.

\nameS starts each run by executing the instrumented
harness with the selected test case to obtain
a firmware trace described in Sec.~\ref{sec:harness}.
The harness identifies the {\em BB context} of
each visited BB using Fuzzware's NVIC implementation
and our call-stack monitoring patches that support coroutines.
Also, it logs the ARM special registers used
for IRQ handling, preventing \name's DSE from
executing ISRs at illegal times.
Finally, it ignores a DTA sink that is a RAM read if
no non-memory instruction in the same BB context
uses the associated sources, as the read then represents
data movement rather than data use.

The model-generation program loads
each firmware binary into \texttt{angr},
which provides the static CFG and all-pairs
shortest paths required in Sec.~\ref{sec:useexec}.
It organizes each firmware trace into lists of
events grouped by BB context to efficiently update
the BB and MMIO pointers in Sec.~\ref{sec:useexec}.
It also hashes the MMIO-read groups obtained
in Sec.~\ref{sec:cluster} to speed up
the consolidation of identical groups.
If a group is matched with firmware strings
not explored in previous runs, the program
short-circuits the DSE on that group and
outputs MMIO models that describe only those 
strings, similar to Redqueen \cite{redqueen}.
Otherwise, the DSE times out after an hour.

Our DSE emulates floating-point operations
slowly when firmware implements them using
integer and bit-vector arithmetic.
Many such operations are functions from
\texttt{math.h} and \texttt{aeabi.h}
with firmware-independent I/O relations.
So, we build Python equivalents that express these
I/O relations directly in floating-point arithmetic.
The DSE speeds up by executing these equivalents
instead of the original functions.

\section{Evaluation}
\label{sec:eval}


We evaluate our \nameS implementation to
address three research questions:
\begin{enumerate}
  \item[\textbf{Q1.}] Does \nameS identify MMIO data chunks
  received by the firmware under test and model each chunk
  across different usage contexts?
  \item[\textbf{Q2.}] Does \nameS enable a given
  firmware fuzzer to achieve higher code coverage
  and trigger more bugs using comparable
  computational resources?
  \item[\textbf{Q3.}] Can \name's DSE still generate
  MMIO models without the heuristics introduced
  in Sec.~\ref{sec:heuristics}?
\end{enumerate}
To answer these questions, we applied \nameS and
vanilla Fuzzware to 24 real-world firmware samples
widely used for benchmarking firmware fuzzers
\cite{pretender, p2im, uEmu,
fuzzware, splits, hoedur, multifuzz}.
Both fuzzers ran on a 64-core
Intel Xeon E5-2683 v4 machine (128GB RAM).
They fuzzed each firmware on four cores and
built MMIO models concurrently on additional cores
in three independent 24-hour trials.

\begin{table}[hbtp]
\caption{MMIO data chunks modeled by \name}
\centering
\begin{tabular}{|l|l|l|}
\hline
\textbf{Firmware} &
\textbf{MMIO Register}&
\textbf{Chunk or Context [Data]}
\\ 
\hline
CNC & RCC PLLCFGR & 
system clock frequency \\ 
& & [PLL factors] \\
\cline{2-3}
 & USART DR & 
G-code [chars] \\ 
\cline{2-3}
 & GPIO IDR & 
debounce sample \\ 
 & & [button + switch states] \\
\cdashline{3-3}
 & & debounce state \\
 & & [debounce samples] \\
\hline
$\bullet$ ChibiOS RTC & 
USART DR & 
command [chars] \\ 
$\bullet$ Console &
UART DR & 
$\ast$ general-purpose \\
\cdashline{3-3}
$\bullet$ Contiki &
UART DR & 
$\ast$ IO/LAN/USB-specific \\
\cdashline{3-3}
$\bullet$ RIOT TWR &
USART RDR & 
+ subcommand \\
\cdashline{3-3}
$\bullet$ Zephyr &
USART RDR & 
+ arguments \\
$\ast$ uTasker &
USART3 DR & 
\\
\hline
$\bullet$ Contiki &
SMWDTHROSC &
elapsed time \\
& ST0$\sim$ST3 & [timer values] \\
$\bullet$ RIOT TWR &
TIM CNT & 
$\bullet$ since startup \\
$\bullet$ Thermostat &
TMR COUNT32 & 
$\ast$ before relock \\
$\ast$ RF Door Lock &
TMR COUNT32 & 
\\
\hline
Contiki &
GPIO MIS + IRQ & 
pin event buffer \\
& DETECT ACK 
& [button ups/downs]\\
\hline
$\ast$ Drone &
I2C DR & 
Accel/Gyro/Mag/Temp
\\ 
$\bullet$ Robot &
I2C DR & 
[8-bit (axis) data] \\ 
\cdashline{3-3}
& & $\ast$ telemetry + PID control \\ 
& & [Accels + Gyros + Mags] \\
\hline
$\bullet$ GPS Tracker &
USART RHR & 
startup AT response \\ 
$\bullet$ LiteOS IoT &
USART RDR & 
[chars + bytes] \\
\cdashline{3-3}
 & & runtime: prefix \\
\cdashline{3-3}
 & & runtime: + payload \\
\hline
Gateway & USART DR & 
Firmata message [bytes] \\ 
\hline
$\bullet$ Heat Press &
UART RHR & 
MODBUS frame \\ 
$\bullet$ PLC &
USART DR & 
$\bullet$ RTU [bytes] \\
$\bullet$ $\ast$ uTasker &
USART2/1 DR &
$\ast$ ASCII [chars] \\
\hline
RF Door Lock &
UART & 
``OK\textbackslash r\textbackslash n'' [chars] \\ 
\cdashline{3-3}
 & TX/RX FIFO &
set/verify password [bytes] \\ 
\hline
RIOT TWR & SPI DR & 
DW1000 device ID [bytes] \\ 
\hline
Steering Control & UART RHR & 
``${<}$servo${>}$,\textbackslash n'' [chars] \\ 
\cdashline{3-3}
 & & ``${<}$servo${>}$,${<}$angle${>}$\textbackslash n'' \\
\hline
XML Parser & 
USART DR & 
infer encoding [XML bytes] \\ 
\hline
Zephyr &
CAN RI0R & 
CAN frame [ID + RTR bits] \\ 
\hline
uTasker USB &
USB OTG FS & 
USB packet [bytes] \\ 
 & DFIFO & \\
\hline
\multicolumn{3}{c}{Contiki: Contiki-NG Shell;
Zephyr: Zephyr SocketCAN;}\\
\multicolumn{3}{c}{uTasker: uTasker MODBUS + uTasker USB}
\end{tabular}
\label{tab:24hrdata}
\end{table}

\subsection{MMIO Data Chunks}
\label{sec:datachunks}

Table~\ref{tab:24hrdata} summarizes the MMIO data
chunks modeled by \nameS and their usage contexts
across these trials.
For each chunk, contexts are separated by dashed lines
and ordered by when \nameS modeled them within a trial.
The table specifies the type of constituent MMIO data
in each chunk, which remains consistent across contexts
unless noted otherwise.
It groups firmware samples with similar chunks
or contexts and marks firmware-specific variations
with $\bullet$ and $\ast$.
Overall, Table~\ref{tab:24hrdata} shows that
\nameS can model diverse MMIO data chunks:
peripheral commands and responses, protocol frames,
software clocks, sensor readings, and pin activities.

Let us analyze \name's identification and modeling of
MMIO data chunks in Table~\ref{tab:24hrdata}.
\nameS identified each chunk by grouping its
constituent data (see Sec.~\ref{sec:cluster}).
Most grouped MMIO data fall into two cases:
(G1) the firmware stored such data in an
array or struct and consumed them later in batches;
(G2) the firmware continuously updates
its state variables with the latest such data.
In both cases, the data in each group formed a complete
data chunk, thus matching the MMIO data of interest
described in Sec.~\ref{sec:motivation}.
We found G1 in peripheral commands, responses,
and protocol frames, G2 in software clocks, and a
mixture of both in sensor readings and pin activities.
For example, Drone retrieves each sensor reading via
six MMIO reads (G1) and adjusts PID control based on 
the latest Accel, Gyro, and Mag readings (G2).
Hence, \nameS identified Drone's chunks adaptively
by first grouping MMIO data into G1 chunks
and then G1 chunks into G2 chunks.

\begin{figure*}[hbtp]
\centerline{
\includegraphics[width=0.9\textwidth]
{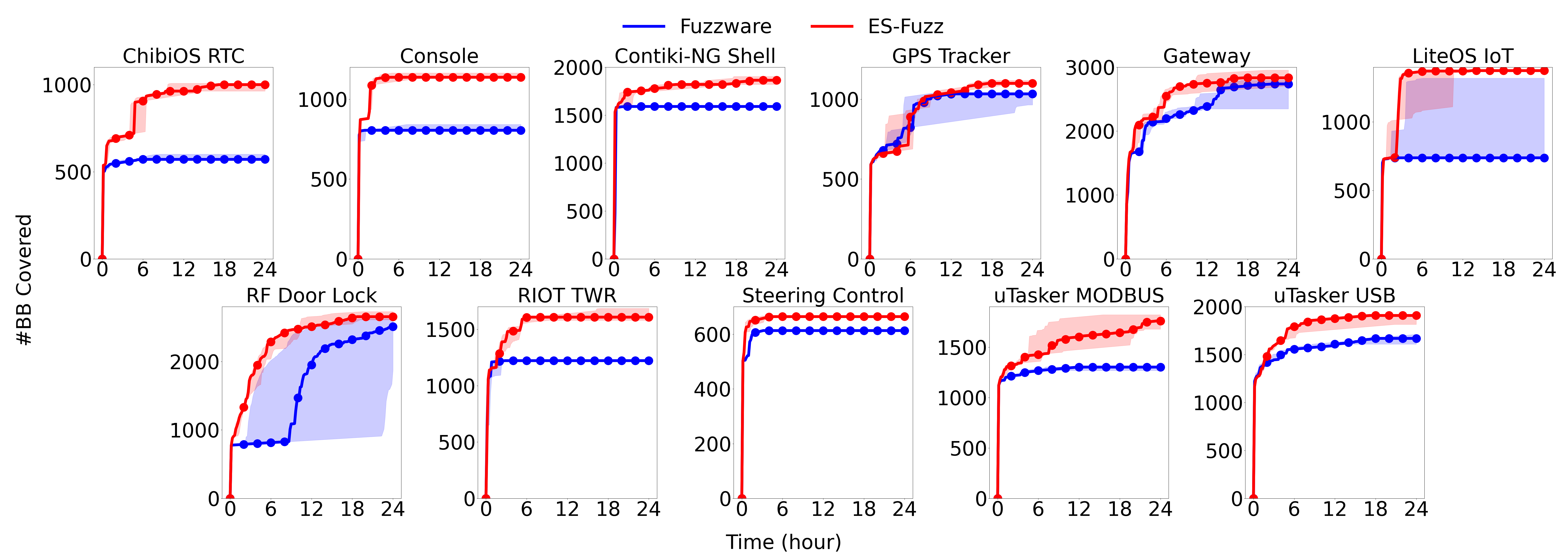}}
\caption{\name's and Fuzzware's coverage in
the firmware under test over 24 hours}
\label{fig:24hrcovs}
\end{figure*}

While most MMIO data groups capture
a firmware-consumed data chunk, a few
contain semantically independent data or subgroups.
ChibiOS RTC updates its software clock by copying
the latest RTC DR, TR, and SSR values instead of
accumulating the time elapsed between updates,
as other firmware does.
\nameS still grouped each RTC register's values
throughout firmware execution into a single chunk:
its DTA operated at byte granularity,
while ChibiOS RTC copied these values at bit granularity.
CNC extracts each G-code from the USART Rx buffer by scanning
unprocessed characters until the first syntax violation.
\nameS thus grouped all characters scanned during
each extraction: those of the current G-code
and that syntax-violating character.
The latter was typically the first character
of the next G-code and further grouped with
characters from the next extraction.
As a result, \nameS correctly grouped MMIO data
for each G-code but then wrongly merged these groups
into a single gigantic chunk.

\name's identification missed a few chunks
since its DTA failed to treat their
constituent MMIO data as sources.
Reflow Oven retrieves each temperature reading
bit-by-bit and immediately converts each bit
to a constant 0/1 via a branch instruction.
The DTA ignored this instruction
and thus lost track of each such bit.
3D Printer would retrieve G-codes in its application code.
However, Fuzzware remained stuck in 3D Printer's
startup code across all trials and hence did not leave
any opportunity for \nameS to identify these G-codes.
This problem relates to the modeling of
{\em status registers}, as discussed in prior 
work \cite{emberio, hoedur, multifuzz}.

\nameS modeled each identified chunk with
MMIO data values that unlocked more code coverage
(see Secs.~\ref {sec:heuristics} and \ref{sec:format}).
Table~\ref{tab:24hrdata} shows that during a trial,
as \nameS completed more runs, its models described
the firmware's expectations for a chunk
more accurately and across more usage contexts.
\nameS modeled the USART3 DR data in uTasker
MODBUS and USB first as main-menu commands,
then as service-specific commands, and
finally as command arguments.
It modeled the USART RHR data in GPS Tracker
as AT responses: first, startup responses;
then, the prefixes of recurring responses;
finally, response payloads.
This adaptive modeling was driven by a cycle:
as Fuzzware encountered new coverage bottlenecks,
\nameS generated new MMIO models, which, in turn,
exposed more code and bottlenecks to Fuzzware.

While \nameS modeled most identified chunks
and most models improved Fuzzware's coverage,
some chunks were missed and some models were ineffective.
The unmodeled chunks fall into two cases:
(M1) modeling them would not unlock additional code coverage;
(M2) \name's DSE for them exhausted
the allocated time or memory before reaching
any unvisited firmware BBs.
M1 typically occurred when the firmware
updated the same state variable with
each MMIO data in the chunk.
Fuzzware could already cover every chunk-dependent BB
via input mutations if the update consistently took
the same execution path and the variable was much
smaller than the chunk.
M2 stemmed from unfaithful DSE or path explosion,
often triggered when \texttt{angr} emulated
conditional instructions in the firmware
(e.g., Robot and Thermostat).
We have already reported these issues
to \texttt{angr} developers.
The ineffective models typically arose from
unfaithful DSE or mismatches between
the modeled MMIO data and firmware literals.
These models did not degrade \name's coverage
across trials since the dummy models
in Sec.~\ref{sec:format} provided a fallback.

\subsection{Code Coverage}
\label{sec:coverage}


Fig.~\ref{fig:24hrcovs} and Table~\ref{tab:24hrcovs}
summarize the code coverage of \nameS and vanilla Fuzzware
across trials.
Fig.~\ref{fig:24hrcovs} focuses on the $11$ firmware
samples where \nameS boosted Fuzzware's coverage.
For each sample, the red and blue curves show
\name's and Fuzzware's median coverage across trials
over 24 hours, with the shaded regions spanning
the minimum and maximum coverage.
Table~\ref{tab:24hrcovs} reports the number of BBs
in each sample and the final coverage of each fuzzer
in individual trials (min/med/max)
and across all trials (total).
Overall, \nameS raised Fuzzware's coverage during a trial
by up to $195\%$ (Fig.~\ref{fig:24hrcovs}, RF Door Lock,
after 8.7 hours) and final coverage across trials by
up to $68\%$ (Table~\ref{tab:24hrcovs}, ChibiOS RTC).

When testing Gateway and RF Door Lock,
\nameS consistently boosted Fuzzware's
coverage in the early stage of every trial
and total coverage across trials.
However, in some trials, \nameS did not
improve Fuzzware's final coverage.
In Gateway, \nameS covered multiple branches in
the Firmata parser by modeling various Firmata messages.
However, its DSE could not model beyond three bytes
per message within the CPU and RAM budgets.
Moreover, the parser is branch-heavy,
taking a branch for each incoming message byte.
AFL's input mutations thus had a chance to compete
with \name's DSE in reaching unvisited BBs.
In RF Door Lock, \nameS gained most of its
additional coverage by modeling the magic word
``OK\textbackslash r\textbackslash n''.
This word guards much of the firmware code, but it is
so short that AFL's input mutations got it by the end
of every trial.

We classify the firmware samples where
\nameS did not improve Fuzzware's coverage
into two cases.
The eight samples at the bottom of Table~\ref{tab:24hrcovs}
are those where \nameS generated no MMIO models.
In such cases, \nameS and vanilla Fuzzware tested firmware
the same way, making coverage comparisons uninformative.
In Sec.~\ref{sec:datachunks} we explained the cases
of 3D Printer, Reflow Oven, Robot, and Thermostat.
Soldering Iron receives data chunks from peripherals
via DMA rather than via MMIO.

The five samples in the middle
of Table~\ref{tab:24hrcovs} are those
for which \name's MMIO models did not improve
Fuzzware's coverage by a noticeable amount.
In these cases, Fuzzware could fall back on
its native models via the dummy models
in Sec.~\ref{sec:format}.
So, \nameS still achieved coverage comparable
to vanilla Fuzzware (${\le}3.5\%$ drop
in the \textbf{total} fields).
We found that \name's models for these samples
did capture the MMIO data chunks listed
in Table~\ref{tab:24hrdata}.
However, deploying these models may have constrained
firmware executions too early in the fuzzing process,
thereby limiting AFL's input mutations.

\subsection{Firmware Vulnerabilities}
\label{sec:bugs}

\nameS enabled Fuzzware to trigger
additional bugs in five firmware samples.
Below we describe these bugs and explain how
\nameS contributed to their discovery.
Some have also been reported by concurrent
works \cite{icicle, splits, multifuzz}.
We are currently in contact with the firmware developers 
regarding the bugs that have not yet been reported.

\textbf{Console.} An out-of-bounds (OOB) read
occurs when the firmware calls the function \texttt{dow}
with a month value greater than 12.
This function helps parse MMIO-retrieved user commands
of the form ``\texttt{rtc \{settime, setalarm\}
YYYY-MM-DD HH:mm:SS}''.
To trigger this bug, Fuzzware must feed the firmware
the command prefixes ``\texttt{rtc \{settime, setalarm\}}''.
This requirement was met during the trials using
\name's MMIO models that describe these prefixes.
Fuzzware did not assign its native MMIO models
to the command-retrieving reads.

\textbf{Contiki-NG Shell.} A null-pointer dereference
occurs when the firmware calls \texttt{strcmp}
to parse a missing argument in
the MMIO-retrieved ``\texttt{rpl-set-root}''.
The firmware assigns a null pointer to this argument
but proceeds with parsing regardless.
To trigger this bug, Fuzzware must feed
``\texttt{rpl-set-root}'' to the firmware
without misspellings.
This requirement was met during the trials
using \name's MMIO model that describes this command.
In contrast, Fuzzware's native MMIO model for
the command-retrieving reads simply extracts
the lowest byte from each 4-byte value
returned by such reads.

\textbf{GPS Tracker.} A null-pointer dereference
occurs when the firmware calls \texttt{strtok}
in \texttt{gsm\_get\_imei} after failing to find
``AT+GSN\textbackslash r\textbackslash r\textbackslash n''
in the received MMIO data stream.
To extract an AT response, the firmware
assumes that the response immediately
follows the corresponding command in the stream.
So, it first calls \texttt{strstr} to locate the command
and then \texttt{strtok} to extract the response.
If the command is absent, \texttt{strstr} returns
a null pointer that breaks \texttt{strtok}.
To trigger this bug, Fuzzware must
progress beyond \texttt{gsm\_set\_pin}
and \texttt{gsm\_wait\_modem\_ready}
in \texttt{gsm\_config}.
This requirement was met during the trials using
\name's MMIO models that describe the required AT responses.
Again, Fuzzware's native MMIO model for
the response-retrieving reads simply extracts
the lowest byte from each 4-byte value
returned by such reads.

\textbf{uTasker USB.} An OOB write occurs when
the firmware calls \texttt{control\_callback} after
extracting an invalidly large interface index from
a \texttt{SET\_LINE\_CODING} request retrieved by MMIO reads.
The firmware copies the request contents into
the USB CDC settings specified by that index,
even when the index exceeds USB\_CDC\_COUNT.
To trigger this bug, Fuzzware must feed the correct
request prefix to the firmware.
This requirement was met during the trials using
\name's MMIO models that describe USB OTG Full-Speed
device data (in bytes).
Fuzzware did not assign its native MMIO models
to the request-retrieving reads.

\textbf{CNC.} The firmware keeps an unaligned
stack pointer (SP) throughout execution after
explicitly initializing the SP to 0x2001FFFF
in \texttt{Reset\_Handler}.
The SP is never 4-byte aligned, and
the firmware will crash on real-world
hardware that requires aligned RAM accesses.
We discovered this bug while diagnosing
a crash in \name's DSE.
Fuzzware tested the firmware smoothly,
as QEMU tolerated the unaligned SP,
but the resulting firmware traces broke \texttt{angr}.
So, we patched Fuzzware's harness \textit{ad hoc}
to keep the SP standard-compliant across firmware.

Recently, FirmReBugger \cite{firmrebugger}
proposes a systematic and automated way to evaluate
firmware fuzzers on a wide range of known bugs.
We plan to extend our evaluation with its framework
once its codebase has been fully updated.

\begin{table*}[hbtp]
\caption{Comparisons between \nameS and vanilla Fuzzware and\\
between our DSE and vanilla DSE in resource efficiency
(O: Yes, X: No)}
\centering
\begin{tabular}{l|r|rr||rr|rr|cc}
\hline
\textbf{Firmware} &
\multicolumn{1}{c|}{\textbf{CPU}} &
\multicolumn{2}{c||}{\textbf{\#BB Covered}} &
\multicolumn{2}{c|}{\textbf{CPU Time}} &
\multicolumn{2}{c|}{\textbf{RAM} (GB)} &
\multicolumn{2}{c}{\textbf{New Model?}} \\
 &
 \multicolumn{1}{c|}{\textbf{Time}} &
 \multicolumn{1}{c}{Fuzzware} &
 \multicolumn{1}{c||}{\name} &
 \multicolumn{1}{c}{vanilla} &
 \multicolumn{1}{c|}{ours} &
 \multicolumn{1}{c}{vanilla} &
 \multicolumn{1}{c|}{ours} &
 \multicolumn{1}{c}{vanilla} &
 \multicolumn{1}{c}{ours} \\
\hline
ChibiOS RTC & 97h & 569 & 963 &
01m 50s & 02m 45s & 1.41 & 1.41 & O & O \\
Console & 202h & 844 & 1164 &
1h 05m 28s & 45s & 52.10 & 0.93 & O & O \\
Contiki-NG Shell & 312h & 1597 & 1820 &
1h 28m 37s & 04m 18s & ${>}$90 & 2.53 & X & O \\
GPS Tracker & 144h & 1023 & 1099 &
27m 13s & 02m 14s & 18.79 & 2.38 & O & O \\
Gateway & 112h & 2558 & 2955 &
04m 39s & 36m 16s & 4.28 & 7.22 & O & O \\
LiteOS IoT & 107h & 1317 & 1372 &
33m 10s & 17m 35s & ${>}$90 & 1.66 & X & O \\
RF Door Lock & 100h & 2236 & 2655 &
1h 21m 19s & 53s & ${>}$90 & 1.49 & X & O \\
RIOT TWR & 110h & 1223 & 1608 &
1h 19m 47s & 08m 26s & ${>}$90 & 2.48 & X & O \\
Steering Control & 122h & 637 & 664 &
2h 44m 14s & 13m 16s & ${>}$90 & 3.14 & X & O \\
uTasker MODBUS & 197h & 1311 & 1678 &
43m 44s & 06m 43s & ${>}$90 & 2.08 & X & O \\
uTasker USB & 115h & 1693 & 1922 &
06m 01s & 06m 10s & 2.24 & 2.24 & O & O \\
\hline
\end{tabular}

\label{tab:resources}
\end{table*}

\subsection{Development and Runtime Overheads}
\label{sec:overheads}

Equipping a given firmware fuzzer with \nameS
incurs additional development effort and runtime overhead.
The development effort stems from adapting
the fuzzer's harness to produce the firmware traces
described in Sec.~\ref{sec:harness} and support
the MMIO models described in Sec.~\ref{sec:format}.
This corresponds to the first part of our
\nameS implementation in Sec.~\ref{sec:implementation}.
This part must be rebuilt if Fuzzware is replaced by
another firmware fuzzer, such as Hoedur \cite{hoedur},
MultiFuzz \cite{multifuzz}, or Khost \cite{khost}.
In contrast, the second part of
our implementation is reusable.
Note that we do not need to adapt
the first part to different firmware under test.

The runtime overhead refers to \name's
CPU and RAM usage for generating adaptive models
of MMIO data chunks, on top of the underlying
fuzzer's resource usage.
To justify this overhead, we compared the code coverage
of our \nameS implementation and vanilla Fuzzware within
the same CPU time.
We ran both fuzzers on the 11 firmware samples
highlighted in Fig.~\ref{fig:24hrcovs} using
the settings in Sec.~\ref{sec:coverage}, except
that vanilla Fuzzware kept running until it spent
as much CPU time as \name.
Both fuzzers generate MMIO models sporadically,
concurrently with fuzzing, and on an indefinite
number of cores.
So, we measured their CPU time using
the CPU accounting option of \texttt{systemd-run}.
The left half of Table~\ref{tab:resources} reports
the CPU time (in hours) and code coverage
of these two fuzzers.
Within the same CPU time, \nameS enabled Fuzzware
to cover more code in every firmware sample.

\subsection{Ablation Study: DSE Heuristics}
\label{sec:efficiency}

To evaluate the heuristics in Sec.~\ref{sec:heuristics} 
for mitigating path explosion, we compared
our DSE with a vanilla DSE: a copy of ours
in which these heuristics were disabled.
Both were evaluated on the 11 firmware samples 
highlighted in Fig.~\ref{fig:24hrcovs}.
We reran a single-core DSE instance per sample
that took the median time to reach an unvisited BB
among all instances in \name's trials of testing that sample.
Each run terminated when the DSE hit an unvisited BB,
aborted when the DSE consumeed ${>}90$GB of RAM,
and never timed out.
Thus, it either succeeded by reaching an unvisited BB
or failed due to memory exhaustion.
The right half of Table~\ref{tab:resources} reports
the CPU times, RAM usage (measured by
\texttt{/usr/bin/time -v}), and outcomes
of our DSE and the vanilla DSE on each sample.

The vanilla DSE failed to reach any
unvisited BB in 6 samples, even with 90GB of RAM.
In Console and GPS Tracker, it required
${86}\times$ and ${11}\times$ more CPU time
as well as ${55}\times$ and ${7}\times$ more RAM,
respectively, to succeed.
While it outperformed our DSE in ChibiOS RTC,
Gateway, and uTasker USB, the resource gaps were
much smaller than those in the samples where
our DSE performed better.
The above observations suggest that a vanilla DSE (e.g.,
used in Laelaps \cite{laelaps} or DevFuzz \cite{devfuzz})
does not directly work for \name's research problem.

In summary, \nameS raised Fuzzware's code coverage
by up to $68$\% in 11 of the 24 firmware samples,
triggered additional bugs in 5 of them, and did not
sacrifice the coverage elsewhere.
Its MMIO models accurately captured a wide range
of data chunks that firmware receives through
multiple MMIO reads across different execution stages.
Moreover, \nameS and its DSE used computational
resources more effectively than vanilla Fuzzware and DSE.

\section{Discussion}
\label{sec:discussions}

There are several ways of improving \name.
First, its MMIO models do not capture the relationships
among values returned from different MMIO registers.
As a result, these models fail to correlate the fields
of a Zephyr CAN frame or the values read from
two GPIO registers for a Contiki button event.
This limitation stems from our design in which \nameS
only groups MMIO reads from the same MMIO register
and in BB contexts of the same IRQ type.
This design keeps our DSE efficient and
our MMIO models easy to deploy.
Several prior works \cite{p2im, uEmu, fuzzware}
make a similar design choice.
One possible extension is to relax
the grouping constraint so that a group may
contain reads from multiple registers of
the same peripheral.

Second, the test case analyzed by \nameS in a run
may not prompt the firmware under test to perform
enough MMIO reads to retrieve a complete data chunk.
In this case, the firmware will not execute much of
its chunk-usage code, thus limiting \name's grouping
of MMIO reads and DSE on each group.
One possible mitigation is to analyze multiple test cases
in a run that, together, cover most firmware BBs reached
by the fuzzer.
However, this will require substantial CPU time and memory.
FIDO \cite{fido} may offer an alternative solution
since it addresses a research problem similar to
the above limitation.

Third, \name's DSE may not replay every ISR along
the path recorded in the given firmware trace.
It executes all required ISRs as soon as possible,
while the fuzzer executes them periodically.
An ISR, when invoked at different times, may take
different paths if its control flow depends on
the firmware's state variables, such as a mutex
and a UART Rx buffer.
Our \nameS implementation mitigates such divergence
by tolerating minor deviations from the recorded path,
postponing an ISR execution, or skipping an ISR.

Fourth, \texttt{angr} may not faithfully
emulate every ARM Thumb instruction.
In our evaluation, it ignored the \texttt{MRS}
and \texttt{MSR} instructions that access
the Interrupt Program Status Register,
stalled on certain conditional instructions
in Robot and Thermostat, and took an infeasible
branch in ChibiOS RTC.
We have already reported these issues to \texttt{angr} 
developers and resolved some with \textit{ad hoc} patches.
One possible extension for \nameS is
to support additional DSE engines.

Lastly, delaying path explosion in DSE
is a longstanding research problem.
In our evaluation, \name's DSE still suffered from
path explosion in some runs despite its overall
advantage over a vanilla DSE in generating
MMIO data-chunk models (see Sec.~\ref{sec:efficiency}).
One possible extension is to restrict each DSE instance
to the call chains identified in Sec.~\ref{sec:cluster},
along which the firmware under test consumes
the modeled MMIO data chunk.

\section{Related Work}
\label{sec:related}

\subsection{Rehosting-based Firmware Fuzzing}
\label{sec:firmfuzz}

Rehosting-based firmware fuzzers test
embedded-system firmware by fuzzing monolithic
firmware binaries in a fully emulated environment.
They typically assume no access to real peripherals,
native OS support, firmware source code, or specifications.
To still achieve decent coverage, they emulate the MMIO,
IRQ \cite{pretender, aim, sEmu, aidfuzzer, fido}, or
DMA \cite{dice, sEmu, gdma} behavior of absent peripherals.
Most such fuzzers rely on MMIO models
to emulate peripheral behavior.

PRETENDER \cite{pretender} models correlated
MMIO reads by replaying peripheral responses
recorded from real-world firmware executions.
P2IM \cite{p2im} classifies each MMIO access by its
surrounding firmware code and models it accordingly.
$\mu$Emu \cite{uEmu} runs DSE on each executed but
unmodeled MMIO read to determine the returned value.
It then reuses this value for the same type of MMIO reads.
Both P2IM and $\mu$Emu avoid modeling
reads from data registers.
In each firmware execution, $\mu$Emu may force
the initial reads from a data register to
retrieve every firmware string once.
Jetset \cite{jetset} models MMIO reads
in the firmware's startup code using DSE
and requires users to specify the DSE destinations.
SEmu \cite{sEmu} automatically extracts MMIO models
from peripheral specifications.
Fuzzware \cite{fuzzware} assigns \underline{fixed}
multi-valued models to \underline{individual} MMIO reads.
It achieves higher code coverage than
the aforementioned works and represents
the SOTA in MMIO modeling.
So, we use it as the underlying
firmware fuzzer to implement \name.
\nameS builds adaptive models of the MMIO data chunks
received by firmware during fuzzing.
As shown in Sec.~\ref{sec:eval}, these models
improve the code and bug coverage of
the underlying firmware fuzzer.

Hoedur \cite{hoedur} and MultiFuzz \cite{multifuzz}
are SOTA firmware fuzzers that redesign their core
fuzzing engines to produce multi-stream firmware inputs.
As noted in these papers, their approaches are
orthogonal and complementary to MMIO modeling.
Hoedur's implementation integrates Fuzzware to
improve coverage using MMIO models, implying that
Hoedur and MultiFuzz could similarly benefit from \name.
Integrating \nameS with either fuzzer requires a one-time
manual revision of the fuzzer's harness.
SplITS \cite{splits} solves a similar problem to
Hoedur's and MultiFuzz's, but specifically for
string comparisons during fuzzing.

Recent studies in MMIO emulation for firmware fuzzing
go beyond MMIO models and multi-stream inputs.
Laelaps \cite{laelaps} runs DSE on every MMIO read
during firmware testing without caching the results.
The DSE spans only a few BBs but
significantly slows down the testing.
Ember-IO \cite{emberio} emulates firmware MMIO
via code instrumentation rather than MMIO models.
Icicle \cite{icicle} is an emulator that provides
firmware fuzzers with an interface to ISA-agnostic
code instrumentation.
SAFIREFUZZ \cite{safirefuzz} and Khost \cite{khost}
rehost Cortex-M firmware on high-performance
ARM platforms for near-native testing.
FirmReBugger \cite{firmrebugger} evaluates
rehosting-based fuzzers on a diverse set of firmware bugs.

\subsection{Structured-input Fuzzing beyond Firmware}
\label{sec:statefuzz}

Fuzzing of structured or stateful inputs has been
applied to programs beyond embedded-system firmware.
Ijon \cite{ijon} leverages human annotations and
inspections during fuzzing to explore the deep
state space of a program.
DevFuzz \cite{devfuzz} navigates the probing logic
of a device driver (analogous to firmware startup code)
using a stateful but deterministic peripheral model.
It builds this model by running a vanilla DSE on
the probing logic and terminating the DSE based on
kernel dynamic debugging.
Morphuzz \cite{morphuzz} identifies
{\em semantic dependencies} in virtual devices
based on the principles of hypervisor design.
ViDeZZo \cite{videzzo} fuzzes virtual devices
using {\em intra-message dependencies}.
A human security analyst is assumed to extract
such dependencies from the source code and
encode them in ViDeZZo's descriptive grammar.
Truman \cite{truman} automatically learns
{\em intra-message, inter-message, and state
dependencies} in the fuzzed virtual devices
from the source code of corresponding OS drivers.
It statically analyzes the drivers by assuming
their use of specific kernel functions.

These works do not address our research problem.
They rely on human annotations, source code, or
native OS support to learn input structures.
Such prerequisites are typically unavailable
in rehosting-based firmware fuzzing and are not
assumed by recent firmware fuzzers, including \nameS
and \cite{p2im, uEmu, fuzzware, hoedur, multifuzz}.

\section{Conclusion}
\label{sec:conclusions}

We have proposed \nameS to build adaptive models
of MMIO data chunks for rehosting-based firmware fuzzing.
It describes MMIO reads in embedded-system firmware
that collectively retrieve a data chunk more accurately
than existing MMIO modeling approaches.
Thus, a SOTA firmware fuzzer can achieve higher
code coverage with \name, which runs
alongside the fuzzer and starts a new run when
the coverage stagnates.
In each run, \nameS attempts to build new MMIO
data-chunk models that unlock additional coverage
in three steps.
First, it produces a detailed trace of the firmware
under test executing a high-coverage test case.
Second, it groups MMIO reads in the trace that
have together retrieved a data chunk.
Third, it models the values returned by the reads
in each group using DSE and heuristics.
The fuzzer will use these MMIO models in
subsequent firmware executions during fuzzing.
We have implemented \nameS on Fuzzware and evaluated it
on 24 firmware benchmarks widely used in prior work.
\nameS is shown to raise Fuzzware's
code coverage by up to $68\%$ in 11 of them,
discover additional vulnerabilities in 5 of them,
and model a wide range of MMIO data chunks.

\section{Fuzzer-specific Heuristics}
\label{sec:taintless}

\nameS models only MMIO reads in the firmware under test
that collectively retrieve a data chunk.
So, based on the firmware fuzzer in use,
the taint analysis in Sec.~\ref{sec:harness} should not
treat the values returned by certain MMIO reads as sources.
If the fuzzer is Fuzzware \cite{fuzzware}, the analysis
should ignore the reads assigned a Constant (single-valued)
or Passthrough (normal-memory) model by the fuzzer.
``Constant'' is typically assigned to reads that
retrieve peripheral status or simple data.
``Passthrough'' is typically assigned to reads
from control registers (see Sec.~\ref{sec:interaction}).
If the fuzzer is $\mu$Emu \cite{uEmu} or P2IM \cite{p2im},
the analysis should ignore the reads not from
the fuzzer-identified data registers.
These reads are likely from status (SR) or control (CR)
registers (see Sec.~\ref{sec:interaction}).
Existing fuzzers emulate such reads well
without \name's MMIO models, by
assigning SR reads with fixed models and
treating CR reads as RAM reads.

\section{Comparison with DevFuzz}
\label{sec:devfuzz}

One may wonder if DevFuzz \cite{devfuzz},
a prior work on fuzz-testing device drivers,
has solved the research problem addressed by \name.
DevFuzz uses DSE to build part of
its MMIO models for driver fuzzing.
\name, Fuzzware \cite{fuzzware}, uEmu \cite{uEmu},
and Jetset \cite{jetset} also use DSE to build
MMIO models for firmware fuzzing.
However, DevFuzz does not address \name's research problem:
its vanilla DSE is prone to path explosion, and
its MMIO-model generation does not support IRQ handling.

Most MMIO modeling approaches for firmware fuzzing
customize their DSE for efficiency.
\name, Fuzzware, uEmu, and Jetset are among them.
Their designs incorporate substantial mechanisms
beyond a vanilla DSE to mitigate path explosion
(see Sec.~\ref{sec:heuristics}).
These mechanisms include well-defined DSE scopes,
heuristics for prioritizing or pruning symbolic
firmware states, a limited set of MMIO reads
symbolized per DSE instance, etc.
\name's DSE heuristics have been evaluated against
path explosion in Sec.~\ref{sec:efficiency}.
The results in the right half of Table~\ref{tab:resources}
show that a vanilla DSE is unlikely to generate
\name's MMIO models before exhausting the allocated memory.

In contrast, DevFuzz does not elaborate on
modifications to its vanilla DSE.
It uses DSE to build part of its MMIO models
(i.e., the {\em probe models}) and encounters
path explosion even in this limited use case.
Sec.~4.1 of \cite{devfuzz} concludes that
``DevFuzz may not be able to generate a probe model.
The reason is that symbolic execution may fail
to complete the probing phase if the probing logic
is too complex to solve within some time budget,
and/or if it requires DMA/IRQ that are rare
so DevFuzz's (current) symbolic execution
does not support.''
DevFuzz mentions its only DSE optimization
in Sec.~4.1 as a stub script that terminates
DSE based on kernel dynamic debugging.
This is far from sufficient compared to
recent firmware fuzzers.
Moreover, a kernel-dependent method does not
apply to rehosting-based firmware fuzzing.

Any solution to \name's research problem must
not only constrain its DSE carefully but also
support IRQ handling during MMIO-model generation.
A large fraction of the MMIO reads modeled by \nameS
occur in ISRs: embedded systems often receive
data chunks spanning multiple MMIO reads
via asynchronous serial communications.
The support for IRQ handling is thus essential.
\name's DSE can execute ISRs for generating MMIO models
(Sec.~\ref{sec:readexec}).
DevFuzz's DSE, as admitted in the above-quoted
statement \cite{devfuzz}, lacks this feature.
DevFuzz's target MMIOs rarely occur in ISRs,
whereas IRQ handling is an important source
of \name's target MMIOs.
This again shows why DevFuzz cannot address
\name's research problem.


Some may still argue that, beyond probe models,
DevFuzz also builds post-probing MMIO models that
target a device driver's behavior after booting.
Likewise, \name's MMIO models primarily support
testing the firmware code after booting.
However, DevFuzz builds such models using
static value analysis rather than DSE.
This analysis requires the program's source code, which
is unavailable in rehosting-based firmware fuzzing.
Moreover, the resulting models do not distinguish
MMIO reads from the same memory and instruction addresses.
In contrast, \nameS is motivated precisely by
the need for MMIO models that distinguish such reads.

\section{More Evaluation Results}
\label{sec:exteval}

\begin{table*}[hbtp]
\caption{\name's and Fuzzware's final coverage
in the firmware under test}
\centering
\begin{tabular}{l|c|rrrr|rrrrr}
\hline
\textbf{Firmware} &
\textbf{Total} &
\multicolumn{4}{c|}{\textbf{Fuzzware}} &
\multicolumn{5}{c}{\name} \\
& \textbf{\#BB} &
\multicolumn{1}{c}{min} & \multicolumn{1}{c}{med} &
\multicolumn{1}{c}{max} & \multicolumn{1}{c|}{total} &
\multicolumn{1}{c}{min} & \multicolumn{1}{c}{med} &
\multicolumn{1}{c}{max} & \multicolumn{2}{c}{total} \\
\hline
ChibiOS RTC & 
3013 &
569 & 571 & 601 & 606 &
963 & 1000 & 1008 & 1020 & (+68.32\%) \\ 
Console & 
2251 &
805 & 805 & 842 & 844 &
1133 & 1137 & 1164 & 1164 & (+37.91\%) \\ 
Contiki-NG Shell & 
4776 &
1588 & 1590 & 1593 & 1594 &
1820 & 1864 & 1904 & 1913 & (+20.01\%) \\ 
GPS Tracker & 
4194 &
964 & 1033 & 1039 & 1042 &
1089 & 1099 & 1118 & 1119 & (+7.39\%) \\
Gateway & 
4921 &
2348 & 2738 & 2795 & 2876 &
2703 & 2834 & 2955 & 3027 & (+5.25\%) \\
LiteOS IoT & 
2423 &
738 & 738 & 1320 & 1320 &
1372 & 1375 & 1376 & 1381 & (+4.62\%) \\ 
RF Door Lock & 
3320 &
1879 & 2511 & 2626 & 2796 &
2609 & 2655 & 2730 & 2975 & (+6.40\%) \\
RIOT TWR & 
4261 &
1222 & 1222 & 1222 & 1222 &
1580 & 1608 & 1686 & 1686 & (+37.97\%) \\ 
Steering Control & 
1835 &
613 & 613 & 615 & 615 &
661 & 664 & 667 & 667 & (+8.46\%) \\ 
uTasker MODBUS & 
3780 &
1268 & 1302 & 1304 & 1304 &
1678 & 1759 & 1821 & 2176 & (+66.87\%) \\ 
uTasker USB & 
3491 &
1610 & 1670 & 1711 & 1739 &
1816 & 1909 & 1922 & 1976 & (+13.63\%) \\
\hline
CNC & 
3614 &
2455 & 2556 & 2573 & 2618 &
2444 & 2522 & 2523 & 2573 & (-1.72\%) \\
Heat Press & 
1837 &
554 & 554 & 554 & 554 &
545 & 545 & 554 & 554 & (+0.00\%) \\
PLC & 
2303 &
626 & 632 & 645 & 645 &
618 & 640 & 640 & 640 & (-0.78\%) \\
XML Parser & 
9376 &
3184 & 3372 & 3822 & 3964 &
3196 & 3439 & 3630 & 3829 & (-3.41\%) \\
Zephyr SocketCAN & 
5943 &
2672 & 2685 & 2691 & 2713 &
2595 & 2622 & 2629 & 2649 & (-2.39\%) \\
\hline
3D Printer &
8045 &
948 & 972 & 1273 & 1313 &
\multicolumn{5}{c}{} \\
6LoWPAN Receiver & 
6977 &
1697 & 2334 & 3197 & 3206 &
\multicolumn{5}{c}{} \\
6LoWPAN Sender & 
6980 &
2239 & 2586 & 3110 & 3195 &
\multicolumn{5}{c}{} \\
Drone &
2728 &
1836 & 1838 & 1841 & 1844 &
\multicolumn{5}{c}{} \\
Reflow Oven &
2947 &
890 & 1188 & 1192 & 1192 &
\multicolumn{5}{c}{} \\
Robot &
3034 &
1295 & 1309 & 1345 & 1351 &
\multicolumn{5}{c}{} \\
Soldering Iron &
3656 &
2321 & 2364 & 2501 & 2523 &
\multicolumn{5}{c}{} \\
Thermostat &
4673 &
2393 & 2944 & 3325 & 3501 &
\multicolumn{5}{c}{} \\
\cline{1-6}
\end{tabular}

\label{tab:24hrcovs}
\end{table*}

We evaluated \nameS with 24 real-world
firmware samples widely used in prior work
\cite{pretender, p2im, uEmu, fuzzware, splits}
by measuring the additional code coverage in them
that \nameS brings to
a given firmware fuzzer.
Specifically, we tested each firmware with both
vanilla Fuzzware and our \nameS implementation
in Sec.~\ref{sec:implementation} three times,
each time for 24 hours.
We compared the code coverage of the two fuzzers
across the three trials, as shown in
Fig.~\ref{fig:24hrcovs} and Table~\ref{tab:24hrcovs}.
Fig.~\ref{fig:24hrcovs} plots the coverage growth
of \nameS and Fuzzware over time in the firmware
where \nameS improved Fuzzware's coverage.
Table~\ref{tab:24hrcovs} lists the total number of
BBs in each firmware and the numbers of those
covered by \nameS and Fuzzware in each trial.
The horizontal lines divide Table~\ref{tab:24hrcovs}
into three parts.
The first part shows the firmware where \nameS
improved Fuzzware's coverage.
The second part shows the firmware where \nameS
generated new MMIO models for Fuzzware but
did not improve Fuzzware's coverage.
The last part shows the firmware where \nameS
did not generate any MMIO models for Fuzzware.
We also compared our \nameS implementation
with vanilla Fuzzware, and \name's DSE with
a vanilla DSE, regarding their resource efficiency.
The results in Table~\ref{tab:resources} justify
\name's use of both additional cores for generating
MMIO models and DSE heuristics in Sec.~\ref{sec:heuristics}.

\end{document}